\documentstyle[a4,12pt]{article}

\newcommand{\beq}{\begin{equation}}
\newcommand{\eeq}{\end{equation}}
\newcommand{\ba}{\begin{eqnarray}}
\newcommand{\ea}{\end{eqnarray}}
\newcommand{\bi}{\begin{itemize}}
\newcommand{\ei}{\end{itemize}}
\newcommand{\bra}{\langle}
\newcommand{\ket}{\rangle}

\def\ps@headings{}
\def\@chapapp{}
\begin{document}
\begin{titlepage}
\vspace{0.2in} 
\begin{flushright}
MITH-98/8  \\ 
\end{flushright} 
\vspace*{1.5cm}
\begin{center} {\LARGE \bf Bose-Einstein vs. electrodynamic condensates:
the question of order and coherence.\\} 
\vspace*{0.8cm}
{\bf  E.~Del Giudice} and {\bf  G.~Preparata} \\
{\it Dept. Physics and INFN-via Celoria, 16-20133 MILAN  \\ 
ITALY }
\\ \vspace*{1cm}
\end{center} 

\begin{abstract}
The remarkable recent experiments on ensembles of magnetically trap\-ped
ultracold alkali  atoms have demonstrated the transition to a highly ordered
phase, that has been attributed to the  process of quantum-mechani\-cal
condensation, predicted long ago (1924) by Bose and Einstein. After having
presented our argument against the above attribution, we show that the known 
phenomenology, including a discrepancy of about one order of magnitude with the
predictions of Bose- Einstein condensation, is in good agreement with the
electromagnetic coherence induced on the alkali  atoms by the long range
electrodynamic interactions. We also predict that for temperatures lower than 
a well defined
$T_{BEC}$, the state predicted by Bose and Einstein coexists with the new
coherent electrodynamic state.
\end{abstract}

\baselineskip=12pt 
\vfill \begin{flushleft}  July 1998 \\
\end{flushleft} 
\end{titlepage}
\baselineskip=12pt

Bose-Einstein ($BE$) condensation, since its theoretical discovery
\cite{1},\cite{2}, has had
a major role in  elucidating subtle aspects of Quantum Field Theory (QFT).  In
particular it has provided a particularly  simple and illuminating realiza\-tion
of the mechanism by which the simplest (and most idealized) of all  many-body
systems, the perfect gas, obeys Nernst's third principle of thermodynamics,
which requires  that entropy go to zero {\em with continuity} as the temperature
tends towards absolute zero.  A property that  the classical, Maxwell-Boltzmann
gas obviously doesn't possess.

But as far as its realization in Nature is
concerned, up until 1995, no firm, generally accepted  evidence was found in
any low temperature physical system.  Well known is the controversy as to 
whether He-II superfluidity is a (particularly striking) manifestation of $BE$
condensation, a possibility  keenly adversed by Landau, and dialectically dealt
with by several other authors \cite{3},\cite{4}.  Be that as it  may, even ignoring the
subtle theoretical problems which we shall take up below, the matter was 
definitely obscured by the fact that He-II and the low-temperature
superconductors are dense  interacting systems, and one can thus legitimately
doubt that the non-interacting perfect quantum gas  be an adequate dynamical
model for them. In order to firmly establish the existence of $BE$-condensation
one thus needs, everybody agrees,  to study highly diluted atomic systems, and
from the condition for $BE$ condensation \footnote{ We use natural units where
$\hbar=c=k_{B}=1$} [$n=\left(\frac{N}{V} \right)$   is the  number density] 
\ba
n_{BE}=\left( \frac{m T_{BE}}{2\pi} \right)^{\frac{3}{2}}  2.612=\nu_{BE} 
\left(\frac{T_{BE}}{\mu K} \right)^{\frac{3}{2}} 
\ea
for a density $n\cong 10^{14} cm^{-3}$ of Na-atoms of mass number $A=23$, at an
average distance $n^{-\frac{1}{3}} \cong 2000 \AA$ (thus practically
non-interacting), one computes $ T_{BE}\cong 2.5~ \mu K   $ an exceedingly
low-temperature.  One  had thus to await for the de\-ve\-lop\-ment of 
sophisticated trapping and cooling techniques to be able to  explore a region
of phase-space where one could unambiguously probe the predictions of Bose and 
Einstein.  And when the temperatures and densities that could be reached were
in the above ball park  punctually the sudden transition to a new state was
observed, whose structure corresponded to the $BE$  condensate \cite{5}.

The transition, in truth, was observed at densities about one order of
ma\-gni\-tu\-de lower than  predicted, but this didn't ring any particular bell
for, according to the authors, it could be accounted for  by experimental
inaccuracies: thus the announcement to the world that $BE$ condensation, more
than  seventy years after its prediction, was an experimental reality.  And the
further observations at MIT \cite{6}  completely confirmed such findings and
their theoretical interpretation. But this remarkable research program,
progressing at a very swift pace, had another very big  surprise in store: at
the beginning of 1997 the same MIT group reported the observation \cite{7} of
strong  interference fringes when two such condensates were let to diffuse one
through the other, and their  measured period
$\Delta=30~\mu$  just
corresponded to the quantum mechanical prediction 
\ba
\Delta =\frac{2\pi}{|\vec{P}|},
\ea
where $P$  is the relative momentum of the two condensates.  Please note that,
according to (2), the  relative velocity of the condensates is in fact quite
small: $v \cong 0.06~ cm/sec$.

The reason why the discovery of interference fringes with the period
(2) is an extremely  significant step forward in our knowledge of the physics
of macroscopic systems of ultracold atoms is  that it is the first experimental
proof that these systems, at least in the phase space region probed by the  MIT
group, are {\em coherent}, i.e. are described by a 
{\em macroscopic wave-function} (or
"order parameter") 
\ba
\Psi \left( \vec{x},t \right)=\bra \Omega |\Psi\left( \vec{x},t  \right)
|\Omega\ket
\ea
where $|\Omega \ket$  is the state of the quantum field ($V$ is the volume of the system and 
$b_{\vec{k}}$ are the quantum  amplitudes associated to the modes of Fourier
momentum $\vec{k}$~) 
\ba
\Psi \left( \vec{x},t \right)= \frac{1}{\sqrt{V}}
\sum_{\vec{k}}b_{\vec{k}}(t)e^{i\vec{k}\vec{x}},
\ea
that collectively describes the atomic system. The important question now is
whether $BE$ condensation actually predicts that the system of  two
condensates, of momentum $-\frac{\vec{P}}{2}$  and $\frac{\vec{P}}{2}$ 
respectively, is described by the macroscopic wave-function
\ba
\Psi \left( \vec{x},t \right)=\sqrt{n}\left( e^{i\frac{\vec{P}}{2} \vec{x}}
+e^{-i\frac{\vec{P}}{2} \vec{x} } \right)
\ea
so that its square, the density of the "matter waves", exhibits just the
observed interference fringes.The answer is a very simple and definite no, at
least for the state which the Bose-Einstein perfect gas  falls into below the
condensation temperature.  Indeed, according to theory, such state contains a
well  defined number of atoms, so that we may set 
\ba
|\Omega\ket=|N_{1} \ket_{\frac{\vec{P}}{2} } |N_{2} \ket_{-\frac{\vec{P}}{2} }
\ea                           				      
and sandwiching the field ($4$) between the states ($6$) we obtain zero,
instead of the observed ($5$),  showing that the $BE$-condensate is indeed
{\em ordered} but not {\em coherent}, and that order and coherence are  two basically
independent physical concepts.  The reason for this (seemingly) surprising
result is  transparent: in absence of interactions among atoms, such as assumed
in the perfect gas model and  deemed perfectly reasonable for the actual
condensates due to their very low density,
\footnote
{As emphasized above, at the typical densities probed, $n \cong 10^{14}~
cm^{-3}$, the average interatomic  distance is $a \cong 2000 \AA$, three orders
of magnitude larger than the atomic size.}
there is no  physical mechanism to bring the phases of the atoms, that are
ordered "by default", to "cohere" in a  macroscopic coherent state, described
by the "order parameter" (5).  In other words, unless a long- range interaction
among the atoms holds, that makes it energetically advantageous for the atoms
to  "align and lock" their phases, there is no reason whatsoever for the $BE$
"order" to transmute itself into  the observed "coherence".  And for pretending
that the discovery of "coherence" be just the  experimental proof that the $BE$
condensate is indeed coherent \cite{7} and not that, instead, $BE$ 
condensation is not what is being observed. 

It must be stressed, however, that the above difficulties of $BE$-condensation
are well present to  the minds of the most thoughtful theoreticians, P.W.
Anderson foremost among them \cite{8}, who believe  that $BE$ condensates may
nevertheless become coherent as a consequence of their mutual interactions. 
However, this view has been challenged in Ref. \cite{4}, on the basis of a set
of common sense arguments,  that convincingly reject the notion that in a
macroscopic system "ignorance may create coherence"   It  is only recently that
a few papers have appeared claiming to have some proof or example in support of 
Anderson's conjecture that a $BE$ condensate develops in due time a macroscopic
phase \cite{9}, \cite{10}. The  fact that here one is dealing with some kind of
paralogism or other is revealed by the claim in Ref. \cite{9}  that the phase
appears spontaneously and kinematically, while it requires a definite
interaction with the  observer in Ref. \cite{10}, whose authors do not care to
explain why in Ref. \cite{9} one obtains for free what  they get only after
having generalized the Copenhagen contorsions to a truly macroscopic system, 
leaving unanswered the very reasonable objections of Leggett and Sols \cite{4}.

To corroborate the view of these latter authors we would like to note that if
one assumes, as  everybody does, that $BE$ condensation initially produces a
state like (6), then any perturbation of the  general type:
\ba
H_{SR}=\frac{1}{2}\int d^{3}x \int d^{3}y 
\Psi(\vec{x},t)^{\dag} \Psi(\vec{x},t) V_{SR}(\vec{x} -\vec{y})
\Psi(\vec{y},t)^{\dag} \Psi(\vec{y},t)
\ea
where $V_{SR}(\vec{x} -\vec{y})$  is a short-range interaction potential, will
lead to an evolution equation for the  order  parameter: 
\ba
i \dot{\Psi}(\vec{x},t) =\bra \Omega |[H_{SR},\Psi(\vec{x},t) ]|\Omega \ket
\ea
whose RHS can easily be shown to vanish for a state of the type (6), a result
that can be directly related  to the gauge-invariance of the Hamiltonian (7),
which, as emphasized time and again by Anderson  himself, prevents the state
(6) from acquiring a non-trivial (relative) phase.  Thus, the only way for the 
atoms' condensate to develop an "order parameter" (5) is that there exists
extra non-perturbative long- range interactions leading to a new ground state
where the atomic systems are in phase, and are  described by a {\em coherent
superposition } of states of the type (6).  Or, put differently, what has been 
experimentally revealed is not a $BE$-condensate but a peculiar coherent state
[see Eq. (5)~] whose  "raison d'\^etre" can only be found in a long-range
interaction totally extraneous, not only to the simple  perfect gas of Bose and
Einstein, but to modern condensed matter theory as well.  We shall call this 
state Coherent Electrodynamic Condensate (CEC). In this paper we shall
demonstrate that the recent experimental findings can be perfectly  understood
and explained in terms of the QED coherent interactions among the ultracold
alkaline  atoms, whose general theory appears in a recently published book
\cite{11}.

We write the Coherence Equations (CE) for the matter and the e.m. field
amplitudes in a box of  volume $V$, as \cite{11} 
\ba
\nonumber
i \dot{a}=\frac{e}{\omega}\left( \frac{N}{V} \right)^{\frac{1}{2}} 
\frac{1}{(2 \omega)^{\frac{1}{2}}} \bra 0 
| \vec{\epsilon}^{r}_{\vec{k}} \cdot \vec{J}  |
\alpha \ket \alpha^{r}_{\vec{k}} b_{\alpha \vec{k}},\\
\nonumber
i \dot{b}_{\alpha \vec{k}}=\frac{e}{\omega}\left( \frac{N}{V}\right)^
{\frac{1}{2}} \frac{1}{(2 \omega)^{\frac{1}{2}}}  \bra \alpha
|{\vec{\epsilon}^{\star r}}_{\vec{k}} \cdot \vec{J}  | 0 \ket  
\alpha^{r}_{\vec{k}} a ,\\
-\frac{1}{2} \ddot { \alpha^{r}_{\vec{k}}}+i \dot { \alpha^{r}_{\vec{k}}}=
\frac{e}{\omega}\left( \frac{N}{V}\right)^{\frac{1}{2}}
\frac{1}{(2 \omega)^{\frac{1}{2}}}
\bra 0 | {\vec{\epsilon}^{\star r} }_{\vec{k}} \cdot \vec{J}  | \alpha \ket 
b_{\alpha \vec{k}} a ,
\ea
where $a$  is the (coherent) amplitude of the matter (atom) field in the ground
state, $b_{\alpha \vec{k}}$   the amplitude  for the excited state $| a \ket$
and
Fourier momentum $\vec{k}$ , $\omega$ the frequency of the transition with
matrix element $ \bra 0 |\vec{J} \cdot  \vec{\epsilon}^{r}_{\vec{k}}| \alpha
\ket$, and $ \alpha^{r}_{\vec{k}} $  the e.m. amplitude of momentum 
$|\vec{k}|=\omega$, and polarisation $r$ with  $\vec{k} \cdot
\vec{\epsilon}^{r}_{\vec{k}}=0$. The time-derivative is with respect to the
adimensional time  $\tau=\omega t$. By restricting the system (9) to a 
"Coherence Domain" (CD), a cubic region of side equal to the wavelength
$\lambda=\frac{2\pi}{\omega}$, for each of the 6  independent Fourier
components $\vec{k}=\omega[(1,0,0),(-1,0,0),...]$, the last two equations of
the system (9)  become: 
\ba
\nonumber
i \dot{b}_{\alpha \vec{k}}=g  {\alpha_{\vec{k}}^{\star}}^{r} a ,\\
-\frac{1}{2} \ddot { \alpha^{r}_{\vec{k}}}+i \dot { \alpha^{r}_{\vec{k}}}=
g b_{\alpha \vec{k}} a,
\ea             
where the coupling constant is 
\ba
g^{2}=\frac{e^{2}}{m_{e}^{2}}\left(\frac{N}{V} \right) \frac{1}{4 \omega^{2}} f,
\ea
and $f$  is the oscillator strength for the transition 
$0 \leftrightarrow \alpha$ .
The very low value of $\left(\frac{N}{V} \right)$ , and the 
correspondingly low value of the coupling constant g clearly imply that the
coherent process involving  the e.m. field modes and the atomic transitions
with frequency $\omega$ is {\em weak}, i.e. the coherent amplitudes
$ \alpha^{r}_{\vec{k}}$ and $  b_{\alpha \vec{k}}$ are of 
$O \left( \frac{1} {N_{CD}^{\frac{1}{2}} } \right)$, where $N_{CD}$  is the number
of atomic systems to be found (in the  average) in the
Coherence Domain.  This latter fact puts an obvious constraint on the density
necessary  for the development of QED coherence in the macroscopic system,
namely: 
\ba
n \geq n_{CE}=\frac{1}{x_{c} \lambda^{3}} =\frac{n_{0}}{x_{c}}
\ea
where $x_{c} (0 \leq x_{c} \leq 1)$ is the fraction of atomic systems involved 
in the coherent process.  The  constraint (12) then simply means that one must
have (in the average) at least {\em one} coherent system per  CD.  By solving
the system (9) -(10) with the appropriate boundary conditions \cite{11} for $n
> n_{0}$  one obtains the following expression for the "gap" associated to  the
Coherent Electrodynamic Condensate  (CEC): 
\ba
\delta=\frac{3 \alpha}{8 \pi^{2}} \frac{\omega^{2}}{m_{e}} f x_{c} =
\delta_{0} x_{c}
\ea						
Please note that $\delta_{0}$  refers to the maximum value of the gap, which,
if one takes into account the  necessary spatial variation \footnote {See
ref. \cite{11}, Ch. 3. },                                   
implies an average value $\bar{\delta_{0}} \cong \frac{1}{2} \delta_{0}$. One
should also notice the linearity  of $\delta$  in $x_{c}$, stemming from the
collective, many-body nature of the coherent e.m. interaction.

Eqs. (12) and (13) embody the main physical features of the CEC, which can now
be brought to  bear upon the recent experiments.  In Table I we report the
relevant parameters for the three alkali  atoms that have been studied so
far \footnote{ Observations of condensates of Li-atoms have been reported in
Ref. \cite{12}} and we add also a line with our prediction for atomic
hydrogen.\\
~\\
TABLE I.{\small The main parameters of the CEC for different atomic systems}
\begin{center}
\begin{tabular}{|c|c|c||c||c||c||c||c|} \hline
Atom & $\lambda(A)$ & $ \omega (eV)$ &$f$ &$n_{0} ~cm^{-3}$ &
$\delta_{0}(\mu K)$ & $\nu_{BE} ~\frac {cm^{-3}}{(\mu K)^{\frac{3}{2}}}$ &
$T_{BEC} (\mu K)$  \\ \hline
Na & $ 5889.9 $ &  $2.13 $ & $.64  $ & $4.89~ 10^{12} $ & $18.7  $ & 
$5.62~10^{13}$ & $0.20$ \\
~ & $  5895.5 $ &  $ 2.13 $ & $ .32 $ & $ 4.87~ 10^{12} $ & $9.35 $ & ~ & ~
\\ \hline
Li & $ 6707   $ &  $1.87 $ & $.50  $ & $3.31~ 10^{12} $ & $11.25 $ & 
$9.40~10^{12}$ & $0.49$  \\ 
~ & $  6707   $ &  $ 1.87 $ & $ .25 $ & $ 3.31~ 10^{12} $ & $5.623$ & ~ & ~
\\ \hline
Rb & $ 7800  $ &  $1.61 $ & $.67  $ & $2.11~ 10^{12} $ & $11.17 $ & 
$4.02~10^{14}$ & $0.03$ \\ 
~ & $  7947   $ &  $ 1.58 $ & $ .33 $ & $ 2.12~ 10^{12} $ & $5.58 $ & ~ & ~
\\ \hline
H  & $ 1215.67$ &  $10.34 $ & $.44  $ & $5.57~ 10^{14} $ & $200  $ & 
$5.08~10^{11}$ & $105$ \\ \hline
\end{tabular}
\end{center}

According to coherent QED \cite{11} for $T \neq 0$  the CEC consists of
two fluids, the coherent one, whose  fraction is $x_c$ , and the incoherent one
which lives in the interstices of the Coherence Domains (CD)  whose fraction is
given by ($m$ is the mass of the atom): 
\ba
(1-x_{c} )=\frac{1}{n} \left( \frac{m T}{2 \pi} \right)^{\frac{3}{2}} f 
\left( \frac{\bar{\delta}}{T} x_{c} \right)
\ea
where 
\ba
f(x)=\frac{2}{\sqrt{\pi}} \int_{0}^{\infty} dt
\frac{t^{\frac{1}{2}}}{(e^{t+x}-1)}
\ea
is a well-known function directly related to the sum over states of a Bose gas. 
The condition for $BE$  condensation can be read off also from (14) once we
put $x_{c}=0$  and recall that $f(0)=2.612$ , thus  obtaining Eq. (1).

In the case of a CEC the condition is rather different, we must first of all
take for the density $n$   the value $n_{CE}$  given by (12), and determine
$x_{c}$   at the transition through:
\ba
\frac{1-x_{c}}{x_{c}}=\left( \frac{n_{BE}}{n_{0}} \right) 
\frac{f\left( \frac{\bar{\delta}}{T} x_{c}\right)}{2.612}
\ea 
which clearly holds only when $n>n_{0}$. Thus when $n_{BE} < n_{0}$, and this 
happens when the temperature $T$  is below the value
\ba
T_{BEC}=\frac{2 \pi}{m} \left( \frac{n_{0}}{2.612}\right)^{\frac{2}{3}}
\ea
for $n_{BE}<n<n_{0}$  our theory predicts an {\em incoherent} $BE$-condensate,
in agreement with the theory of  Bose and Einstein.

The existence for $T<T_{BEC}$  of an interval of
densities where the CEC cannot occur, leaving as the  only possibility the
formation of an incoherent $BE$C, is a most significant prediction of our theory, 
which allows the test of the predicted coherence properties of the two
condensates as well as of the  occurrence of a phase transition between them
when  $n=n_{0}$.
 
In Fig. 1 we show our predictions for the phase diagram for an
ensemble of both Na and Rb  atoms.  One sees that below a line, which only in a
limited range of densities is different from the line  given by Eq. (1), we
predict condensation, which exhibits the coherence properties, typical of the 
electrodynamic condensate, only when $n \geq n_{0}=\frac{1}{\lambda^{3}}$,
where $\lambda$ is the wavelength of the line with largest  oscillator 
strength. 

Apart from the prediction that the MIT experiments are in the CEC-region, which
has been  dramatically verified by the observation of strong interference
between two condensates \cite{7}, our phase- diagram predicts a large deviation
from the "kinematics" of $BE$C for the Boulder experiments \cite{5},  while for
MIT \cite{6} no such deviation occurs.  Indeed, a glance at Fig. 1 shows that
when the condensate   first appears in the Boulder experiments, i.e. for
$T=170~nK$  and $n=2.6~10^{12}~cm^{-3}$, we are far away  (almost an order of
magnitude) from $n_{BE}=2.94~10^{13}~cm^{-3}$  predicted by the Bose-Einstein
theory,  while our theory predicts that at $T=170~nK$  the transition density
is about $n_{0}=2.11~10^{12}~cm^{-3}$, in full  agreement with observations. 
On the other hand, in the case of the MIT experiments, where the  transition is
observed for $T \cong 2\mu K$, we predict it to occur for $n \cong n_{BE}$, in
agreement with experiment.Finally we note that, according to our theory, the
maximum temperature $T_{BEC}$  for which a phase- transition between $BE$C and
CEC may be observed is $190~nK$ and $30~nK$  for Na and Rb atoms  respectively. 

In conclusion, we have shown that QED is capable to finally solve the puzzle
raised by the  discovery of coherence in condensates of highly diluted,
ultracold alkali atoms.  Having provided  further arguments against the
likelihood (and reasonableness) of coherence in $BE$C's, we have outlined  the
theoretical bases for the emergence of Coherent Electrodynamic Condensates
(CEC's) and  described their main features.  We have then shown that the phase
diagram is essentially modified,  there appearing for $T<T_{BEC}$  a line of
transition between the $BE$C and the CEC, and in addition  deviations from the
"kinematics" of $BE$C that affect the Boulder \cite{6} but not the MIT \cite{7}
experiments

\newpage
\begin{center}
FIGURE CAPTION
\end{center}
\begin{figure}[h]
\small
\caption{  The phase diagram for the condensation of Rb and Na Atoms. The
lines  are the boundaries of the regions where
the Coherent Electrodynamics Condensation (CEC) occurs.}
\end{figure}

\end{document}